\documentstyle[12pt,epsf,epsfig]{article}
\begin{document}

\title{Magnetic Moment of The $\Theta^+$ Pentaquark State}
\author{Peng-Zhi Huang, Wei-Zhen Deng, Xiao-Lin Chen and Shi-Lin Zhu\\
Department of Physics, Peking University, Beijing 100871, China}
\maketitle

\begin{abstract}

We have calculated the magnetic moment of the recently observed
$\Theta^+$ pentaquark in the framework of the light cone QCD sum
rules using the photon distribution amplitudes. We find that
$\mu_{\Theta^+}=(0.12\pm 0.06) \mu_N$, which is quite small. We
also compare our result with predictions of other groups.

\end{abstract}
\vspace{0.3cm}

\pagenumbering{arabic}

%%%%%%%%%%%%%%%%%%%%%%%%%%%%%%%%%%%%%%%%%%%%%%%%%%
\section{Introduction}\label{sec1}
%%%%%%%%%%%%%%%%%%%%%%%%%%%%%%%%%%%%%%%%%%%%%%%%%%

Early this year LEPS Collaboration at the SPring-8 facility in
Japan observed a sharp resonance $\Theta^+$ at $1.54\pm 0.01$ GeV
with a width smaller than 25 MeV and a statistical significance of
$4.6\sigma$ in the reaction $\gamma n \to K^+ K^- n$ \cite{leps}.
This resonance decays into $K^+ n$, hence carries strangeness
$S=+1$ and $B=+1$.

Later the same resonance was confirmed by several other groups. In
a different reaction $K^+ Xe\to \Theta^+ Xe^\prime \to K^0 p
Xe^\prime$, DIANA Collaboration at ITEP observed this resonance at
$1539\pm 2$ MeV with a width less than $9$ MeV \cite{diana}. Now
$\Theta^+$ decays into $K^0 p$. The confidence level is
$4.4\sigma$.

CLAS Collaboration in Hall B at JLAB observed $\Theta^+$ in the
$K^+ n$ invariant mass spectrum at $1542\pm 5$ MeV in the
exclusive measurement of the $\gamma d \to K^+ K^- pn $ reaction
\cite{clas}. The statistical significance is $5.3\sigma$. The
measured width is $21$ MeV, consistent with CLAS detector
resolution. There was only preliminary hint that $\Theta^+$ might
be an iso-singlet from the featureless $M(K^+ p)$ spectrum in the
CLAS measurement.

SAPHIR Collaboration observed this positive-strangeness resonance
in the $\rm nK^+$ invariant mass distribution with a $4.8\sigma$
confidence level in the photoproduction of the $\rm nK^+K^0_s$
final state with the SAPHIR detector at the Bonn ELectron
Stretcher Accelerator ELSA \cite{saphir}. Its mass is found to be
$\rm M_{\Theta^+} = 1540\pm 4\pm 2$ MeV.  An upper limit of $\rm
\Gamma_{\Theta^+} < 25$ MeV was set for the width of this
resonance at $90\%$ confidence level. From the absence of a signal
in the $\rm pK^+$ invariant mass distribution in $\rm\gamma p\to
pK^+K^-$ at the expected strength they further concluded that the
$\Theta^+$ must be an isoscalar.

Recently another very important observation came from the high
energy collision experiment. NA49 Collaboration found evidence for
the existence of a narrow $\Xi^- \pi^-$ baryon resonance with mass
of $(1.862\pm 0.002) $ GeV and width below the detector resolution
of about 0.018 GeV in proton-proton collisions at $\sqrt{s}=17.2$
GeV \cite{na49}. The quantum number of this state is $Q=-2, S =
-2, I = 3/2$ and its quark content is $(d s d s \bar u)$. They
also observed signals for the $Q=0$ member of the same isospin
quartet with a quark content of $(d s u s \bar d)$ in the $\Xi^-
\pi^+$ spectrum. The corresponding anti-baryon spectra also show
enhancements at the same invariant mass.

Due to the complicated infrared behavior of Quantum Chromodynamics
(QCD), it's nearly impossible to predict the hadron spectrum
analytically from first principle. Lattice simulation may provide
an alternate feasible way to extract the whole spectrum in the
future. But now, people have just been able to understand the
first orbital and radial excitation of the nucleon on the lattice
\cite{liu}.

Experimentally there have accumulated tremendous data in the low
energy sector in the past decades. Under such a circumstance,
various QCD-inspired models were proposed. Among them, the simple
quark model (QM) has been surprisingly successful in the
classification of hadrons and calculation of their spectrum and
other low-energy properties \cite{isgur}. According to QM, mesons
are composed of a pair of quark and anti-quark while baryons are
composed of three quarks. Both mesons and baryons are color
singlets. Nearly all the experimentally observed hadrons fit into
the quark model classification scheme quite nicely.

In contrast, QCD itself does allow the existence of the
non-conventional hadrons with the quark content other than $q\bar
q$ or $q q q$, which is beyond conventional mesons and baryons in
the quark model. Some examples are glueballs ($gg, ggg, \cdots$),
hybrid mesons ($q\bar q g$), and other multi-quark states ($qq\bar
q \bar q$, $qqqq\bar q$, $qqq\bar q \bar q \bar q$, $qqqqqq,
\cdots$). In fact, hybrid mesons are found to mix freely with
conventional mesons in the large $N_c$ limit \cite{tom}. However,
despite extensive experimental searches in the past two decades,
none of these states has been firmly established until this year
\cite{pdg}.

The surprising discovery of very narrow resonance with positive
strangeness by LEPS, DIANA, CLAS, SAPHIR and NA49 Collaboration
shall be a milestone in the hadron spectroscopy, if these states
are further established experimentally. Perhaps, a new landscape
is emerging on the horizon, of which we have only had a first
glimpse through the above experiments. Now arises a natural
question: are there other "genuine" hadrons with valence quark
(anti-quark) number $N=4, 6, 7, 8, \cdots$, in which quarks do not
form two or more color-singlet clusters such as hadronic molecules
or nuclei? Is there an upper limit for $N$? In our universe, there
may exist quark stars where the quark number is huge. Is there a
gap in $N$ from pentaquark states to quark stars? All these are
very interesting issues awaiting further experimental exploration.

On the other hand, these experiments have triggered heated
discussions of the interpretation of these resonances
\cite{zhu,bob,early,others,buddy,cohen,princeton,zhao,mag}. Up to
now, the parity and angular moment of the $\Theta^+$ particle have
not been determined while its isospin has not been cross-checked
by other groups.

The partial motivation of the recent experimental search of the
pentaquark state came from the work by Diakonov et al.
\cite{diak}. They proposed the possible existence of the $S=1$
$J^P={1\over 2}^+$ resonance at $1530$ MeV with a width less than
15 MeV using the chiral soliton model and argued that $\Theta^+$
is the lightest member of the anti-decuplet multiplet which is the
third rotational state of the chiral soliton model (CSM). Assuming
that the $N(1710)$ is a member of the anti-decuplet, $\Theta^+$
mass is fixed with the symmetry consideration of the model.

However, identifying $N(1710)$ as a member of the anti-decuplet in
the CSM is kind of arbitrary \cite{page}. Instead, if the
anti-decuplet $P_{11}$ is $N(1440)$, the $\Theta^+$ would be
stable as the ground state octet with a very low mass while
$\Theta^+$ would be very broad with the anti-decuplet $P_{11}$
being $N(2100)$ \cite{page}. Furthermore, if the decay width of
the anti-decuplet $N(1710)$ was shifted upwards to be comparable
with PDG values, the predicted width of $\Theta^+$ particle would
have exceeded the present experimental upper bound \cite{page}.
Moreover, the mass of the pentaquark state with the quark content
$(d s d s \bar u)$ is rigourously predicted in the CSM to be
$2070$ MeV, which is $210$ MeV higher than the experimental value
measured by NA49 Collaboration \cite{na49}.

A more serious challenge to the chiral soliton model came from the
large $N_c$ consistency consideration by Cohen \cite{cohen}. He
found that predictions for a light collective $\Theta^+$ baryon
state (with strangeness +1) based on the collective quantization
of chiral soliton models are shown to be inconsistent with large
$N_c$ QCD since collective quantization is legitimate only for
excitations which vanish as $N_c \to \infty$. He concluded that
the prediction for $\Theta^+$ properties based on collective
quantization of CSM was not valid \cite{cohen}.

The relationship between the bound state and the SU(3) rigid
rotator approaches to strangeness in the Skyrme model was
investigated in \cite{princeton}. It was found that the exotic
state may be an artifact of the rigid rotator approach to the
Skyrme model for large $N_c$ and small $m_K$.

Jaffe and Wilczek proposed that the observed $\Theta^+$ state
could be composed of an anti-strange quark and two highly
correlated up and down quark pairs arising from strong color-spin
correlation \cite{bob}. The resulting $J^P$ of $\Theta^+$ is
${1\over 2}^+$. They predicted the isospin $3/2$ multiplet of
$\Xi$ $(ddss \bar u)$ with $S=-2$ and$J^{\Pi}={1\over 2}^{+}$
around 1750 MeV. Such a state with the same quantum number was
observed by NA49 but with a much higher mass at $1860$ MeV
\cite{na49}.

We have estimated the mass of the pentaquark state with QCD sum
rules and found that pentaquark states with isospin $I=0, 1, 2$
lie close to each other around $(1.55\pm 0.15)$ GeV. Unfortunately
we are unable to determine its parity. However, we pointed out
that the experimentally observed baryon resonance $\Theta^+
(1540)$ with $S=+1$ can be consistently identified as a pentaquark
state if its $J^P={1\over 2}^-$. Such a state was expected in QCD.
If its parity is positive, this pentaquark state would be really
exotic. We emphasized that the outstanding issue is to determine
its quantum numbers experimentally.

In the present work, we shall employ the light cone QCD sum rules
(LCQSR) to extract the magnetic moment of the $\Theta^+$ particle.
The baryon magnetic moment is another fundamental observable as
its mass, which encodes information of the underlying quark
structure and dynamics. Different models generally predict
different values. Such a study will deepen our knowledge of
pentaquark states and may help us explore its dynamics and
distinguish so many models in the literature.

Our paper is organized as follows: Section I is an introduction. A
brief review of this field is presented. In Section II we
summarize our previous work on the pentaquark mass sum rule. Then
we present the formalism of LCQSR in Section III. Our numerical
analysis and discussions are given in Section IV, where we also
compare our result with other groups' prediction.

%%%%%%%%%%%%%%%%%%%%%%%%%%%%%%%%%%%%%%%%%%%%%%
\section{Mass Sum Rule}\label{sec2}
%%%%%%%%%%%%%%%%%%%%%%%%%%%%%%%%%%%%%%%%%%%%%%

The method of QCD sum rules incorporates two basic properties of
QCD in the low energy domain: confinement and approximate chiral
symmetry and its spontaneous breaking. One considers a correlation
function of some specific interpolating currents with the proper
quantum numbers and calculates the correlator perturbatively
starting from high energy region. Then the resonance region is
approached where non-perturbative corrections in terms of various
condensates gradually become important. Using the operator product
expansion, the spectral density of the correlator at the quark
gluon level can be obtained in QCD. On the other hand, the
spectral density can be expressed in term of physical observables
like masses, decay constants, coupling constants etc at the hadron
level. With the assumption of quark hadron duality these two
spectral densities can be related to each other. In this way one
can extract hadron masses etc. For the past two decades QCD sum
rule has proven to be a very powerful and successful
non-perturbative method \cite{svz,reinders}.

Due to the low mass of $\Theta^+$, we have argued its angular
momentum is likely to be one half and considered the correlator
for $I=0$ pentaquark state \cite{zhu}
\begin{equation}\label{cor-1}
i\int d^4 x e^{ipx} \langle 0|T\{ \eta_0 (x), {\bar \eta}_0 (0) \}
|0\rangle\ = \Pi (p) {\hat p} + \Pi' (p )
\end{equation}
where $\bar \eta =\eta^\dag \gamma_0$ and ${\hat p} =p_\mu \cdot
\gamma^\mu$. The interpolating current takes the form \cite{zhu}
\begin{eqnarray}\label{cu3}
\eta_0(x)={1\over \sqrt{2}} \epsilon^{abc} [u^T_a(x) C\gamma_5 d_b
(x)] \{ u_e (x) {\bar s}_e (x) i\gamma_5 d_c(x) - \left(
u\leftrightarrow d\right) \}
\end{eqnarray}
where $a, b, c$ etc are the color indices. $T$ denotes transpose.
$C$ is the charge conjugation matrix. $\left( C\gamma_5\right)^T
=-C\gamma_5$ ensures the isospin of the up and down quark pair
inside the first bracket to be zero. The anti-symmetrization in
the second bracket ensures that the isospin of the other up and
down quark pair is also zero.

The overlapping amplitude $f_0$ of the interpolating current with
the pentaquark state was defined as
\begin{eqnarray}
\label{decay} \langle 0|\eta_0(0)|p, I=0\rangle= f_0 u(p)
\end{eqnarray}
where $u(p)$ is the Dirac spinor of pentaquark field with $I=0$.

At the hadron level, the chiral even structure $\Pi (p)$ can be
expressed as
\begin{equation}
\Pi (p)= {f^2_0 \over p^2 -M_0^2} + \mbox{higher states}
\end{equation}
where $M_0$ is the pentaquark mass. On the other hand, it will be
calculated in terms of quarks and gluons.
\begin{eqnarray}\label{quark}\nonumber
&\langle 0|T\{\eta_0(x) \bar \eta_0(0)\}|0\rangle
=-\epsilon^{abc}\epsilon^{a'b'c'} \times &\nonumber\\ &\{ -{\bf
\mbox{Tr}}\left[iS_d^{bb'}(x) \gamma_5 C iS_u^{Taa'}(x) C \gamma_5
\right]{\bf \mbox{Tr}}\left[ i\gamma_5 iS_d^{cc'}(x)
i\gamma_5 iS_s^{ee'}(-x) \right] iS_u^{ee'}(x) &\nonumber\\
&+{\bf \mbox{Tr}}\left[i\gamma_5 iS_d^{cc'}(x) i\gamma_5
iS_s^{ee'}(-x) \right]iS_u^{ea'}(x) \gamma_5 CiS_d^{Tbb'}(x) C
\gamma_5 iS_u^{ae'}(x)&\nonumber\\
&- {\bf \mbox{Tr}}\left[iS_d^{bb'}(x) i\gamma_5 iS_s^{ee'}(-x)
{i\gamma_5} iS_d^{cc'}(x) \gamma_5 C iS_u^{Taa'}(x)C \gamma_5 \right] iS_u^{ee'}(x) &\nonumber\\
&+iS_u^{ea'}(x) \gamma_5 C\left[iS_d^{bb'}(x) i\gamma_5
iS_s^{ee'}(-x)
i\gamma_5 iS_d^{cc'}(x) \right]^T C \gamma_5iS_u^{ae'}(x)&\nonumber\\
&- {\bf \mbox{Tr}}\left[iS_d^{bb'}(x) \gamma_5 C iS_u^{Taa'}(x) C
\gamma_5 \right]
iS_u^{ec'}(x) i\gamma_5 iS_s^{ee'}(-x) i\gamma_5 iS_d^{ce'}(x) &\nonumber\\
&+ iS_u^{ec'}(x)i\gamma_5iS_s^{ee'}(-x) i\gamma_5 iS_d^{cb'}(x)
\gamma_5 C iS_u^{Taa'}(x) C\gamma_5 iS_d^{be'}(x)&\nonumber\\
&+iS_u^{ea'}(x) \gamma_5 C iS_d^{Tbb'}(x) C \gamma_5
iS_u^{ac'}(x) i\gamma_5 iS_s^{ee'}(-x) i\gamma_5 iS_d^{ce'}(x) &\nonumber\\
&+iS_u^{ea'}(x) \gamma_5 C \left[ iS_u^{ac'}(x) i\gamma_5
iS_s^{ee'}(-x) i\gamma_5 iS_d^{cb'}(x) \right]^T C \gamma_5
iS_d^{be'}(x) \}&
\end{eqnarray}
where $iS_s^{ee'}(-x)$ is the strange quark propagator in the
coordinate space.

After making Fourier transformation to the above equation and
invoking Borel transformation to Eq. (\ref{cor-1}) we have
obtained the mass sum rule \cite{zhu}
\begin{equation}\label{mass}
f^2_0e^{-{M_0^2\over M^2}}=\int_{m_s^2}^{s_0} e^{-{s\over M^2}}
\rho_0 (s) ds
\end{equation}
where $m_s$ is the strange quark mass, $\rho_0 (s)$ is the
spectral density and $s_0$ is the threshold parameter used to
subtract the higher state contribution with the help of
quark-hadron duality assumption. Roughly speaking, $\sqrt{s_0}$ is
around the first radial excitation mass. The spectral density
reads
\begin{eqnarray}\label{spectral}
\rho_0=\frac{1}{(2\pi)^8}[{3 s^5\over  2^8
7!}+{s^2\over96}({5\over12}a_q^2+{11\over
24}a_sa_q)+({7\over432}a_sa_q^3+{1\over 864}a_q^4) \delta (s)]
\end{eqnarray}
where we have used the factorization approximation for the
multi-quark condensates.

The pentaquark mass was found to be \cite{zhu}
\begin{eqnarray}
M_0^2 =\frac{\int_{m_s^2}^{s_0}  e^{-{s/M^2}} \rho' (s) d
s}{\int_{m_s^2}^{s_0}  e^{-{s/M^2}} \rho (s)} d s
\end{eqnarray}
with $\rho'(s)=s\rho(s)$ except that $\rho'(s)$ does not contain
the last term in $\rho(s)$.

In the numerical analysis, we have used the values of various QCD
condensates $a_q =-(2\pi)^2\langle\bar qq\rangle=0.55
\mbox{GeV}^3$, $a_s=0.8a_q=0.44 \mbox{GeV}^3$. We used $m_s (1
\mbox{GeV}) =0.15$ GeV for the strange quark mass in the ${\bar
{MS}}$ scheme. Numerically we arrived at $ M_0 =(1.56\pm 0.15)
\mbox{GeV}$, where the central value corresponds to $M^2=2$
GeV$^2$ and $s_0=4$ GeV$^2$.

%%%%%%%%%%%%%%%%%%%%%%%%%%%%%%%%%%%%%%%%%%%%%%%%%%%%%
\section{Formalism of Light Cone QCD Sum Rules}\label{sec3}
%%%%%%%%%%%%%%%%%%%%%%%%%%%%%%%%%%%%%%%%%%%%%%%%%%%%%

The LCQSR is quite different from the conventional mass QSR, which
is based on the short-distance operator product expansion (OPE).
The LCQSR is based on the OPE on the light cone, which is the
expansion over the twists of the operators \cite{bbk}. The main
contribution comes from the lowest twist operator. Matrix elements
of nonlocal operators sandwiched between a photon (or hadronic
state) and the vacuum define the photon (hadron) distribution
amplitudes. When the LCQSR is used to calculate the coupling
constant, the double Borel transformation is always invoked so
that the excited states and the continuum contribution can be
treated quite nicely. Moreover, the final sum rule depends only on
the value of the photon (or hadron) distribution amplitude at a
specific point, which is much better known than the whole
distribution function. In the present case our sum rule involves
the photon light cone distribution amplitudes
$\varphi_{\gamma}(u_0 ={1\over 2})$. Note this parameter is
universal in all processes at a given scale. In this respect,
$\varphi_{\gamma}(u_0 ={1\over 2})$ is a fundamental quantity like
the quark condensate, which is to be determined with various
non-perturbative methods. Like the quark condensate, it can be
extracted consistently through the analysis of the light cone sum
rules.

In the framework of QCD sum rules, the nucleon magnetic moment was
first studied using the external field method in Refs.
\cite{ioffe,balit,kogan}. The presence of the electromagnetic
field will polarize the vacuum and lead to a few new induced
condensates with various universal vacuum susceptibilities. Later
this formalism was extended to extract the magnetic moments of the
octet and decuplet baryon and heavy baryons \cite{chiu,zsl}.
LCQSRs were first used to calculate the magnetic moments of
nucleons in \cite{first}. Recently the magnetic moments of the
octet and decuplet baryons were reformulated and discussed with
the help of the light cone QCD sum rule technique \cite{aliev}.

In the present case, we are interested in the pentaquark magnetic
moments. We shall consider the following correlator
\begin{eqnarray}
 \Pi (p_1, p_2, q)=  i\int d^4 x \, e^{i p x}
\langle\gamma(q)|T\{\eta_0(x) \bar \eta_0(0) \}|0\rangle
\end{eqnarray}
where $\gamma$ represents the external electromagnetic field with
the vector potential $B_\mu (x)= \varepsilon_\mu e^{iq\cdot x} $.
$\varepsilon_\mu$ is the photon polarization vector. Throughout
this work we shall use the convention of outgoing photons. Its
field strength is $F_{\mu\nu}(x)=(-i)(\varepsilon_\mu q_\nu
-\varepsilon_\nu q_\mu) e^{iq\cdot x}$. $p_1=p, p_2=p_1+q$ is the
final and initial pentaquark momentum.

At the hadron level, the correlator can be expressed as
\begin{eqnarray} \label{pp}\nonumber
\Pi (p_1, p_2, q) =& f_0^2\varepsilon^\mu\frac{{\not
p}_1+m_0}{p_1^2-m_0^2} [F_1(q^2)\gamma_\mu
+\frac{i\sigma_{\mu\nu}q^\nu }{2m_0}F_2(q^2)]\frac{{\not
 p}_2+m_0}{p_2^2-m_0^2}
\\ \nonumber
&+f_0 f_\ast\varepsilon^\mu\frac{{\not p}_1+m_0}{p_1^2-m_0^2}
[F^\ast_1(q^2) \gamma_\mu
+\frac{i\sigma_{\mu\nu}q^\nu}{m_0+m_\ast}F^\ast_2(q^2)]
\frac{{\not p}_2+m_\ast}{p_2^2-m_\ast^2} \\ & + f_\ast f_0
\varepsilon^\mu\frac{{\not p}_1+m_\ast}{p_1^2-m_\ast^2}
[F^\ast_1(q^2) \gamma_\mu
+\frac{i\sigma_{\mu\nu}q^\nu}{m_0+m_\ast}F^\ast_2(q^2)]
\frac{{\not p}_2+m_0}{p_2^2-m_0^2} +\cdots
\end{eqnarray}
where $m_\ast$ is the mass of the excited pentaquark state,
$f_\ast$ is the overlapping amplitude of our interpolating current
with these states. $F^\ast_{1,2}(q^2)$ are the electromagnetic
transition form factors between the ground state and excited
pentaquarks.

We have used the electromagnetic vertex of $\Theta^+$ in writing
down the above formula.
\begin{eqnarray}\label{vertex}
\langle \Theta^+(p_1)|\Theta^+(p_2)\rangle_\gamma
=\varepsilon^\mu\bar{u}_0(p_1)[F_1(q^2)\gamma_\mu
+\frac{i\sigma_{\mu\nu}q^\nu }{2m_0}F_2(q^2)]u_0(p_2)
\end{eqnarray}

In Eq. (\ref{pp}) the first term contains two poles at both
$p^2_1=m_0^2$ and $p_2^2=m_0^2$ with both the initial and final
baryon being the ground state pentaquark.  We have also explicitly
written down terms with a single pole either at $p^2_1=m_0^2$ or
$p_2^2=m_0^2$. In this case one of the initial or final state is
the excited state. The ellipse denotes the continuum contribution.
As we will show below, the contribution from all the non-diagonal
terms in Eq. ({\ref{pp}) will be either eliminated or strongly
suppressed after we invoke double Borel transformation with the
variables $p_1^2, p_2^2$ simultaneously.

Rewriting Eq. (\ref{pp}) we get
\begin{eqnarray}\label{pole}\nonumber
\Pi (p_1, p_2, q)& = f_0^2\varepsilon^\mu\frac{\not
 p_1+m_0}{p_1^2-m_0^2} [(F_1(q^2)+ F_2(q^2)) \gamma_\mu
+\frac{(p_1+p_2)\mu}{2m_0}F_2(q^2)]\frac{\not
 p_2+m_0}{p_2^2-m_0^2} +\cdots \\  &=
\frac{f_0^2}{(p_1^2-m_0^2)(p_2^2-m_0^2)} \left[F_1(q^2)+
F_2(q^2)\right]  \not\!p_1 \not\!\varepsilon \not\!p_2 +\cdots
\end{eqnarray}
The pentaquark magnetic moment is defined as
\begin{equation}
\mu_{\Theta^+}=\left[ F_1(0)+ F_2 (0) \right] {e_{\Theta^+}\over
2m_0}
\end{equation}

We are only interested in the term involving the magnetic form
factor $F_1(q^2)+F_2(q^2)$. So we focus on the tensor structure
$\not\!p_1 \not\varepsilon \not\!p_2$, which is equivalent to
$-i\epsilon_{\mu\nu\alpha\beta}\gamma^\mu\gamma_5\varepsilon^\nu
q^\alpha p^\beta$ up to terms containing a single gamma matrix.

At the quark gluon level, the expression of the above correlator
can be obtained through simple replacement in Eq. (\ref{quark}).
There are two classes of diagrams according to the way how the
photon couples to the quark lines. First, the photon couples to
the quark line perturbatively through the standard QED
interaction. For this set of diagrams, we may replace one of the
free quark propagator in Eq. (\ref{quark}) by the one with the
electromagnetic interaction
\begin{equation}
\langle 0|T\{q^a(x){\bar q}^b(0)\}|0\rangle_{F_{\mu\nu}}
=\frac{\delta^{ab}e_q}{16\pi^2x^2}\int_0^1
du\{2(1-2u)x_\mu\gamma_\nu
+i\epsilon_{\mu\nu\rho\sigma}\gamma_5\gamma^\rho x^\sigma
\}F^{\mu\nu}(ux)
\end{equation}
\begin{equation}
\langle 0|T\{q^a(0){\bar q}^b(x)\}|0\rangle_{F_{\mu\nu}}
=-\frac{\delta^{ab}e_q}{16\pi^2x^2}\int_0^1
du\{2(1-2u)x_\mu\gamma_\nu
+i\epsilon_{\mu\nu\rho\sigma}\gamma_5\gamma^\rho x^\sigma
\}F^{\mu\nu}[(1-u)x]
\end{equation}
where we have adopted the Fock-Schwinger gauge $x^\mu A_\mu(x)=0$
to express the electromagnetic vector potential in terms of the
gauge invariant $F_{\mu\nu}$.

The second class of diagrams involve the non-perturbative
interaction of photons with the quarks in terms of the photon
light cone distribution amplitude. One of the five propagators in
Eq. {(\ref{quark}) is substituted by
\begin{eqnarray}
\langle \gamma (q)|q^a(x){\bar q}^b(0)|0\rangle
=&-\frac{\sigma^{\mu\nu}}{8}\langle \gamma (q)|{\bar q}^b(0)
\sigma_{\mu\nu}q^a(x)|0\rangle
+\frac{\gamma^\mu\gamma_5}{4}\langle
\gamma (q)|{\bar q}^b(0) \gamma_\mu\gamma_5q^a(x)|0\rangle\nonumber\\
 &-\frac{\gamma^\mu}{4}\langle \gamma (q)|{\bar q}^b(0)
\gamma_\mu q^a(x)|0\rangle
\end{eqnarray}

The two-particle photon light cone distribution amplitudes (LCPDA)
are defined as \cite{bbk,photon}:
\begin{eqnarray}
\langle \gamma (q)|\bar q(x) \sigma_{\mu\nu}q(0)|0\rangle =&i e_q
\langle\bar qq\rangle \int_0^1 du \, e^{iu q x}
((\varepsilon_\alpha q_\beta - \varepsilon_\beta q_\alpha) \{ \chi
\varphi(u) + x^2 [(g_1(u) \nonumber\\ &- g_2(u)]\}+ \{qx
(\varepsilon_\alpha x_\beta - \varepsilon_\beta x_\alpha) +
\varepsilon x (x_\alpha q_\beta - x_\beta q_\alpha)\} g_2(u))
\end{eqnarray}
\begin{eqnarray}
\langle \gamma (q)|\bar q(x) \gamma_\mu\gamma_5q(0)|0\rangle =
\frac{f}{4} e_q \epsilon_{\mu\nu\rho\sigma} \varepsilon^\nu q^\rho
x^\sigma \int_0^1 due^{iuqx} \psi(u)
\end{eqnarray}
\begin{eqnarray}
\langle \gamma (q)|\bar q(x) \gamma_\mu q(0)|0\rangle = f^{(V)}
e_q \varepsilon_\mu \int_0^1 due^{i uqx} \psi^{(V)}(u)
\end{eqnarray}
The $\varphi (u)$ is associated with the leading twist-two photon
wave function, while $g_1(u)$ and $g_2(u)$ are twist-4 LCPDAs. All
these LCPDAs are normalized to unity, $\int_0^1 du \; f (u) =1$.
In the above formula, the summation over quark color indices is
implicitly assumed.

With these definitions and spatial translation transformation,
it's easy to derive
\begin{eqnarray}
\langle \gamma (q)|\bar q(0) \sigma_{\mu\nu}q(x)|0\rangle =&i e_q
\langle\bar qq\rangle \int_0^1 du \, e^{i(1- u) q x}
((\varepsilon_\alpha q_\beta - \varepsilon_\beta q_\alpha) \{ \chi
\varphi(u) + x^2 [(g_1(u) \nonumber\\ &- g_2(u)]\} + \{qx
(\varepsilon_\alpha x_\beta - \varepsilon_\beta x_\alpha) +
\varepsilon x (x_\alpha q_\beta - x_\beta q_\alpha)\} g_2(u))
\end{eqnarray}
\begin{eqnarray}
\langle \gamma (q)|\bar q(0) \gamma_\mu\gamma_5q(x)|0\rangle =
-\frac{f}{4} e_q \epsilon_{\mu\nu\rho\sigma} \varepsilon^\nu
q^\rho x^\sigma \int_0^1 due^{i(1- u)qx} \psi(u)
\end{eqnarray}
\begin{eqnarray}
\langle \gamma (q)|\bar q(0) \gamma_\mu q(x)|0\rangle = f^{(V)}
e_q \varepsilon_\mu \int_0^1 due^{i(1- u)qx} \psi^{(V)}(u)
\end{eqnarray}

After tedious but straightforward calculation we arrive at the
correlator in the coordinate space, to which we then make Fourier
transformation. The formulas are:
\begin{equation}\label{ft1}
\int {e^{ipx}\over (x^2)^n} d^D x = i (-1)^{n+1}
{2^{D-2n}\pi^{D/2} \over (-p^2)^{D/2 -n}} {\Gamma (D/2 -n)\over
\Gamma (n)} \; ,
\end{equation}
\begin{equation}\label{ft2}
\int {{\hat x}e^{ipx}\over (x^2)^n} d^D x =  (-1)^{n+1}
{2^{D-2n+1}\pi^{D/2} \over (-p^2)^{D/2+1 -n}} {\Gamma (D/2+1
-n)\over \Gamma (n)} {\hat p} \; .
\end{equation}

After isolating the correct tensor structure, we further make
double Borel transformation with the variables $p_1^2$ and
$p_2^2$. In this way the single-pole terms in (\ref{pole}) are
eliminated. The formula reads:
\begin{equation}\label{double}
{{\cal  B}_1}^{M_1^2}_{p_1^2} {{\cal  B}_2}^{M_2^2}_{p_2^2}
{\Gamma (n)\over [ m^2 -(1-u)p_1^2-up^2_2]^n }= (M^2)^{2-n}
e^{-{m^2\over M^2}} \delta (u-u_0 ) \; .
\end{equation}
Here $M=\frac{M_1^2M_2^2}{M_1^2+M_2^2}$ is the Borel parameter and
$u_0\equiv\frac{M_1^2}{M_1^2+M_2^2},
1-u_0\equiv\frac{M_2^2}{M_1^2+M_2^2}$.

Subtracting the continuum contribution which is modelled by the
dispersion integral in the region $s_1 , s_2\ge s_0$, we arrive
at:
\begin{eqnarray}\label{mag}\nonumber
&(2\pi)^8 f_0^2 |[F_1(0)+F_2(0)]| e^{-\frac{M_0^2}{M^2}}=-\{
\frac{\pi^2}{5!2^4}(14e_u+3e_d) f \psi (1-u_0)
M^{10}f_4(\frac{s_0}{M^2})&\\\nonumber &+\frac{\pi^2}{5!2^3}e_s f
\psi (u_0) M^{10}f_4(\frac{s_0}{M^2})
-\frac{\pi^2}{3^22^6}(14e_u+3e_d) fm_sa_s \psi
(1-u_0)M^{6}f_2(\frac{s_0}{M^2})&\\\nonumber &+\frac{\pi^2}{3^2
2^5}(20e_u+5e_d) fm_sa_q \psi
(1-u_0)M^{6}f_2(\frac{s_0}{M^2})&\\\nonumber &+\frac{\pi^2}{3^2
2^5}(20e_u+5e_d) fa_sa_q \psi (1-u_0)
M^{4}f_1(\frac{s_0}{M^2})&\\\nonumber &+\frac{\pi^2}{3^2
2^5}(20e_u+5e_d) fa_q^2 \psi (1-u_0)
M^{4}f_1(\frac{s_0}{M^2})&\\\nonumber &+\frac{\pi^2}{3^2 2^2}e_s
fa_q^2 \psi (u_0) M^{4}f_1(\frac{s_0}{M^2}) -\frac{1}{5!
2^6}(14e_u+3e_d+2e_s) M^{12}f_5(\frac{s_0}{M^2})&\\\nonumber
&+\frac{1}{3\times2^9}(14e_u+3e_d) m_s
a_sM^{8}f_3(\frac{s_0}{M^2})-\frac{1}{3\times2^8}(20e_u+5e_d) m_s
a_qM^{8}f_3(\frac{s_0}{M^2}) &\\\nonumber
&-\frac{1}{3^22^6}(20e_u+5e_d) a_s
a_qM^{6}f_2(\frac{s_0}{M^2})-\frac{1}{3^2 2^6}(20e_u+5e_d+8e_s)
a_q^2M^{6}f_2(\frac{s_0}{M^2}) &\\\nonumber &-\frac{1}{3^3
2^5}(14e_u+3e_d) a_sa_q^3-\frac{1}{3^3 2^4}e_s a_q^4 +\frac{1}{5!
2^5}(e_u+2e_d) m_sa_q \chi \varphi (1-u_0)
M^{10}f_4(\frac{s_0}{M^2})&\\\nonumber
&-\frac{1}{3\times2^6}(e_u+2e_d) m_sa_q [g_1(1-u_0)-g_2(1-u_0)]
M^{8}f_3(\frac{s_0}{M^2})&\\\nonumber &+\frac{1}{3^2
2^8}(e_u+2e_d) a_sa_q \chi \varphi (1-u_0)
M^{8}f_3(\frac{s_0}{M^2})&\\\nonumber &-\frac{1}{3^2
2^4}(e_u+2e_d) a_sa_q [g_1(1-u_0)-g_2(1-u_0)]
M^{6}f_2(\frac{s_0}{M^2})& \\\nonumber
&-\frac{1}{3\times2^8}(e_u-2e_d) a_q^2 \chi \varphi (1-u_0)
M^{8}f_3(\frac{s_0}{M^2})&\\\nonumber
&+\frac{1}{3\times2^4}(e_u-2e_d) a_q^2 [g_1(1-u_0)-g_2(1-u_0)]
M^{6}f_2(\frac{s_0}{M^2})&\\\nonumber &+\frac{1}{3\times2^7} e_s
a_sa_q \chi \varphi (u_0)M^{8}f_3(\frac{s_0}{M^2})&\\\nonumber
&-\frac{1}{3\times2^3}e_s a_sa_q [g_1(u_0)-g_2(u_0)]
M^{6}f_2(\frac{s_0}{M^2}) -\frac{1}{3^2 2^5}(e_u-2e_d) a_sa_q^3
\chi \varphi (1-u_0) M^{2}f_0(\frac{s_0}{M^2})&\\\nonumber
&+\frac{1}{3^2 2^4}e_s a_sa_q^3 \chi \varphi (u_0)
M^{2}f_0(\frac{s_0}{M^2}) +\frac{1}{3^2 2^3}(e_u-2e_d) a_sa_q^3
[g_1(1-u_0)-g_2(1-u_0)] &\\\nonumber &-\frac{1}{3^2 2^2}e_s
a_sa_q^3 [g_1(u_0)-g_2(u_0)]+\frac{1}{3^3 2^5}(e_u+2e_d) a_q^4
\chi \varphi (1-u_0) M^{2}f_0(\frac{s_0}{M^2})&\\\nonumber &
-\frac{1}{3^3 2^3}(e_u+2e_d) a_q^4
[g_1(1-u_0)-g_2(1-u_0)]&\\\nonumber
&+\frac{1}{3\times2^7}(e_u-2e_d) m_sa_sa_q^2 \chi \varphi (1-u_0)
M^{4}f_1(\frac{s_0}{M^2})&\\\nonumber
&-\frac{1}{3\times2^4}(e_u-2e_d) m_sa_sa_q^2
[g_1(1-u_0)-g_2(1-u_0)] M^{2}f_0(\frac{s_0}{M^2})&\\\nonumber
&-\frac{1}{3 \times 2^6}(e_u-2e_d) m_sa_q^3 \chi \varphi (1-u_0)
M^{4}f_1(\frac{s_0}{M^2})&\\
&+\frac{1}{3\times2^3}(e_u-2e_d) m_sa_q^3 [g_1(1-u_0)-g_2(1-u_0)]
M^{2}f_0(\frac{s_0}{M^2})\}&
\end{eqnarray}
where we have used the functions
$f_n(x)=1-e^{-x}\sum^n_{k=0}\frac{x^k}{k!}$ to subtract the
excited states and continuum contribution.

The left side of the above equation is obtained from double Borel
transformation to the hadron level correlator. We have assumed the
quark-hadron duality. Since in the present case, both the initial
and final states are the same pentaquark. It is natural to employ
$M_1^2=M_2^2=2M^2$ so we have $u_0=\frac{1}{2}$.

%%%%%%%%%%%%%%%%%%%%%%%%%%%%%%%%%%%%%%%%%%%%%%
\section{Results and Discussions}\label{sec4}
%%%%%%%%%%%%%%%%%%%%%%%%%%%%%%%%%%%%%%%%%%%%%%

Dividing our light cone sum rule for the pentaquark magnetic
moment Eq. (\ref{mag}) by its mass sum rule Eq. (\ref{mass}), we
can extract the combination $|F_1(0)+F_2(0)|$. In the numerical
analysis, we follow Refs. \cite{bbk,photon} and use the following
form for the photon light cone distribution amplitudes,
\begin{eqnarray} \nonumber
\psi(u)=1 , & \\ \nonumber  \varphi(u)=6u(1-u) ,& \\ \nonumber
g_1(u)=-\frac{1}{8}(1-u)(3-u) ,& \\ \nonumber
g_2(u)=-\frac{1}{4}(1-u)^2
\end{eqnarray}
with $f=0.028 \mbox{GeV}^2$, $\chi (1 \mbox{GeV}) =-4.4
\mbox{GeV}^{-2}$ \cite{ioffe,balit,kogan,bbk,photon}.

The variation of the pentaquark magnetic moment with $M^2, s_0$ is
shown in Figure 1. Numerically we get
\begin{equation}
|\mu_{\Theta^+}|=(0.20\pm 0.10) {e_{\Theta^+}\over 2m_0} ,
\end{equation}
In unit of nucleon magneton, we have
\begin{equation}
|\mu_{\Theta^+}|=(0.12\pm 0.06) \mu_N ,
\end{equation}

The central value is obtained at $M^2=2.0$GeV$^2$ and $s_0=4.0$
GeV$^2$. The errors come from (1) the uncertainty of the values of
the photon light cone distribution amplitudes at $u_0={1\over 2}$;
(2) the truncation of the expansion over the twist and keeping
only the lowest-twist few terms containing two particles; (3) the
truncation of the operator product expansion in the calculation of
the terms not involving the photon LCPDA; (4) the uncertainty of
the condensate values; (5) the variation of the sum rule with the
continuum threshold and the Borel parameter within the working
interval; (6) the neglect of the higher dimension condensates; (7)
the neglect of perturbative QCD corrections etc.

%---------figure 1-----------------------------
\begin{figure}\label{fig1}
\epsfxsize=12cm \centerline{\epsffile{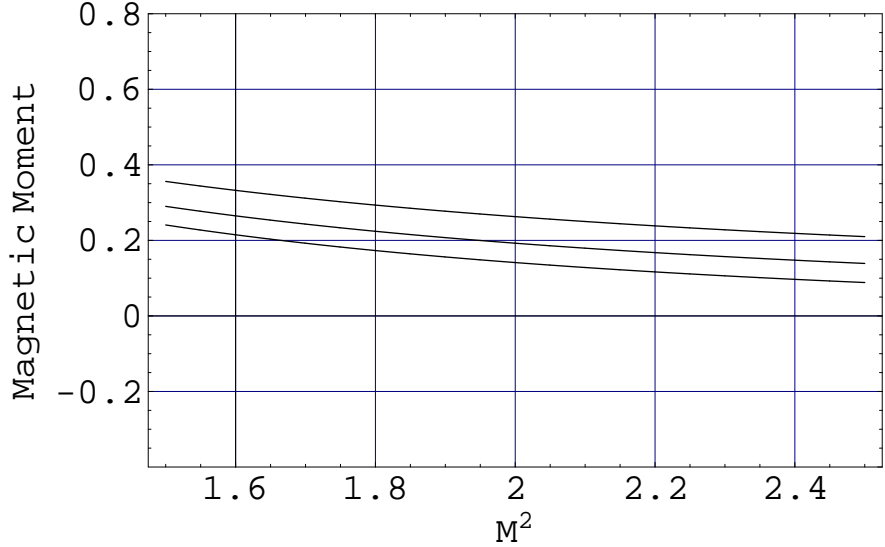}} \vspace{0.5cm}
\caption{The variation of the absolute value of the magnetic
moment of $\Theta^+$ (in unit of ${e_{\Theta^+}\over 2m_0}$), with
the Borel parameter $M^2$ and the continuum threshold $s_0$. The
three curves in this figure (from top to bottom) correspond to
$s_0=3.61, 4.0, 4.41\mbox{GeV}^2$ respectively.}
\end{figure}

In our calculation we have assumed the pentaquark state $\Theta^+$
is an isoscalar with $J={1\over 2}$. No assumption is made of its
parity. In fact, our interpolating current couples to pentaquark
with both negative and positive parity as pointed out in Ref.
\cite{zhu}. Our formalism picks out only the state with lowest
mass without knowledge of its parity. Fortunately the
electromagnetic vertex of pentaquarks with either negative or
positive parity is the same, which ensures that we can extract the
absolute value of the magnetic moment of the lowest pentaquark
state even if we do not know its parity. Our result shows that the
magnetic moment of $\Theta^+$ is quite small.

In Ref. \cite{hosaka},  Nam, Hosaka and Kim first estimated the
anomalous magnetic moment $\kappa_{\Theta^+}$ of the $\Theta^+$
pentaquark when they discussed the photo-production of $\Theta^+$
from the neutron. For example, they have estimated
$\kappa_{\Theta^+} = - 0.7$ and $\kappa_{\Theta^+} = - 0.2$ for
$J^P={1\over 2}^+$ and $J^P={1\over 2}^-$ respectively using Jaffe
and Wilczek's diquark picture. Also, they have considered the
possibility of a K N bound system, with $\kappa(KN, J^P={1\over
2}^+) = - 0.4$ and $\kappa(KN, J^P={1\over 2}^-) =- 0.5$. These
numbers are all in unit of $\Theta^+$ magneton ${e_0\over
2m_{\Theta^+}}$.

In Ref. \cite{zhao}, a quark model calculation of the $\Theta^+$
magnetic moment was performed using Jaffe and Wilczek's picture
for the pentaquark \cite{bob}. For comparison, we borrow the
relevant two formulae from Ref. \cite{zhao} and list them below.
First Zhao wrote down \cite{zhao}:
\begin{equation}
\mu_\theta s_z(s_z={1\over 2})= \langle\Theta^+ \Big|\left[
\mu_{ud}{\bf 0} +\mu_{ud}{\bf 0} + \mu_{\bar{s}} \frac{\bf 1}{\bf
2} +\frac{e_{ud}}{2m_{ud}}{\bf 1}\right]
 \Big|\Theta^+\rangle (s_z={1\over 2}) ,
\end{equation}
with the bold numbers denoting the spin and orbital angular
momentum vector. We note in passing that the author has made an
implicit assumption that $ud$ quark pairs are point-like  and
there is no quark exchange between these two pairs. Otherwise the
above simple formula may not hold.

Then Zhao obtained \cite{zhao}
\begin{equation}
{1\over 2}\mu_\theta ={1\over 2} a^2 \left[\langle 1 \ 0,{1\over
2} {1\over 2} \Big| {1\over 2} {1\over 2}\rangle^2 - \langle 1 \ 1
, {1\over 2} -{1\over 2}\Big| {1\over 2} {1\over 2}\rangle^2
\right] \mu_{\bar{s}} + b^2 \langle 1 \ 1 , {1\over 2} -{1\over
2}\Big| {1\over 2} {1\over 2}\rangle^2 \frac{e_{ud}}{2m_{ud}} ,
\end{equation}
With $\mu_{\bar{s}}\equiv e_{\bar{s}}/2m_s$, $m_s=500$ MeV,
$e_{ud}\equiv e_u+e_d=e_0/3$, $m_{ud}=720$ MeV and assuming equal
probability for two spatial configuration, i.e., $a=b={1\over
\sqrt{2}}$, he finally obtained \cite{zhao}
\begin{equation}
\mu_{\Theta^+}=0.13{e\over 2m_0}
\end{equation}

In Ref. \cite{mag} Kim and Praszalowicz derived relations for the
anti-decuplet within the framework of chiral soliton model in the
chiral limit. The $\Theta^+$ pentaquark magnetic moment is
estimated to be $(0.2\sim 0.3)\mu_N$ \cite{mag}. It's interesting
to note that magnetic moment of $\theta$ pentaquark derived from
different models is also quite small.

In short summary, we have estimated the magnetic moment of the
$\Theta^+$ pentaquark state in the framework of light cone QCD sum
rules using the photon distribution amplitude. All the necessary
parameters in the present calculation have been determined from
previous studies. To this extent, our calculation may be viewed as
a robust prediction.

The width of the $\Theta^+$ pentaquark is so narrow. With the
accumulation of events, its magnetic moment may be extracted from
experiments eventually in the near future, which may help
distinguish different theoretical models and deepen our
understanding of the underlying dynamics governing its formation.

This project was supported by the National Natural Science
Foundation of China under Grant 10375003, Ministry of Education of
China, FANEDD and SRF for ROCS, SEM.

%---------------------------------------------------------------------------


\begin{thebibliography}{99}
\bibitem{leps}T. Nakano et al., Phys. Rev. Lett. 91, 012002
(2003).
\bibitem{diana}V. V. Barmin et al., hep-ex/0304040.
\bibitem{clas}S. Stepanyan et al., hep-ex/0307018.
\bibitem{saphir}J. Barth et al., hep-ph/0307083.
\bibitem{na49}NA49 Collaboration, hep-ex/0310014.
\bibitem{liu}S. J. Dong et al., hep-ph/0306199.
\bibitem{isgur}S. Godfrey and N. Isgur, Phys. Rev. D 32, 189
(1985);S. Capstick and N. Isgur, Phys. Rev. D 34, 2809 (1986).
\bibitem{tom}T. D. Cohen, Phys. Lett. B 427, 348 (1998).
\bibitem{pdg}Particle Dada Group, Phys. Rev. D 66, 010001 (2002).
\bibitem{zhu}Shi-Lin Zhu, hep-ph/0307345, Phys. Rev. Lett. 91, 232002 (2003).
\bibitem{bob}R. Jaffe and F. Wilczek, hep-ph/0307341,
Phys. Rev. Lett. 91, 232003 (2003).

\bibitem{early}H. Gao and B.-Q. Ma, hep-ph/0305294, Mod. Phys. Lett.
A 14, 2313 (1999);M. V. Ployakov and A. Rathke, hep-ph/0303138;F.
Stancu and D. O. Riska, hep-ph/0307010; B. G. Wybourne,
hep-ph/0307170; A. Hosaka, hep-ph/0307232; T. Hyodo, A. Hosaka, E.
Oset, nucl-th/0307105; M. Karliner and H. J. Lipkin,
hep-ph/0307243, hep-ph/0307343.

\bibitem{others} P.V. Pobylitsa, hep-ph/0310221; D. Diakonov, V. Petrov,
hep-ph/0310212; J. Letessier, G. Torrieri, S. Steinke, J.
Rafelski, hep-ph/0310188; I.M.Narodetskii et al., hep-ph/0310118;
N. Auerbach, V. Zelevinsky, nucl-th/0310029; D.E. Kahana, S.H.
Kahana, hep-ph/0310026; J. Haidenbauer, G. Krein, hep-ph/0309243;
D. Diakonov, V. Petrov, hep-ph/0309203;L. Ya. Glozman,
hep-ph/0309092, hep-ph/0308232; M. Praszalowicz, hep-ph/0308114;
R.A. Arndt, I.I. Strakovsky, R.L. Workman, nucl-th/0308012; X.
Chen, Y. Mao, B.-Q. Ma, hep-ph/0307381; S. Nussinov,
hep-ph/0307357; J. Randrup, nucl-th/0307042; M.V. Polyakov, A.
Rathke, hep-ph/0303138; H. Walliser, V.B. Kopeliovich,
hep-ph/0304058.

\bibitem{buddy} R. Bijker, M.M. Giannini, E. Santopinto, hep-ph/0310281;
Y. Oh, H. Kim, S. H. Lee, hep-ph/0310117, hep-ph/0310019,
hep-ph/0311054; R.D. Matheus et al., hep-ph/0309001; F.Huang,
Z.Y.Zhang, Y.W.Yu, B.S.Zou, hep-ph/0310040; C. E. Carlson et al.,
hep-ph/0310038, hep-ph/0307396; J. Sugiyama, T. Doi, M. Oka,
hep-ph/0309271;  F.Csikor, Z. Fodor, S.D. Katz, T.G. Kovacs,
hep-lat/0309090; S. Sasaki, hep-lat/0310014; B. Jennings, K.
Maltman, hep-ph/0308286; K. Cheung, hep-ph/0308176; P.Bicudo, G.
M. Marques, hep-ph/0308073; L. W. Chen, V. Greco, C. M. Ko, S. H.
Lee, W. Liu, nucl-th/0308006;W. Liu, C. M. Ko, V. Kubarovsky,
nucl-th/0310087; W. Liu, C. M. Ko, nucl-th/0309023,
nucl-th/0308034; D. Borisyuk, M. Faber, A. Kobushkin,
hep-ph/0307370; Felipe J. Llanes-Estrada, E. Oset, V. Mateu,
nucl-th/0311020.

\bibitem{cohen}T. D. Cohen, R. F. Lebed, hep-ph/0309150; T. D.
Cohen, hep-ph/0309111.

\bibitem{princeton}N. Itzhaki et al., hep-ph/0309305.

\bibitem{hosaka}S.I.Nam, A.Hosaka, H.-Ch.Kim, hep-ph/0308313.

\bibitem{zhao}Q. Zhao, hep-ph/0310350.

\bibitem{mag}H.-C. Kim and M. Praszalowicz, hep-ph/0308242.

\bibitem{diak}D. Diakonov, V. Petrov, and M. Ployakov, Z. Phys. A
359, 305 (1997).

\bibitem{page}S. Capstick, P. R. Page, and W. Roberts, hep-ph/0307019.

\bibitem{svz}M.A. Shifman, A.I. Vainshtein and V.I. Zakharov, Nucl. Phys. {\bf
B 147}, 385 (1979).

\bibitem{reinders}L. J. Reinders, H. Rubinstein, and S. Yazaki,
Phys. Rept. 127, 1 (1985).

\bibitem{bbk}I. I. Balitsky, V. M. Braun, and A. V. Kolesnichenko,
Nucl. Phys. B 312, 509 (1989).

\bibitem{ioffe}B. L. Ioffe and A. V. Smilga, Nucl. Phys. B {\bf 232}, 109 (1984).

\bibitem{balit}I. I. Balitsky and A. V. Yung, Phys. Lett. B {\bf 129},
328 (1983); I. I. Balitsky and A. V. Kolesnichenko, Yad. Fiz. {\bf
41}, 282 (1985).

\bibitem{kogan}V. M. Belyaev and Ya. I. Kogan, Yad. Fiz {\bf 40}, 1035 (1984).
(Sov. J. Nucl. Phys. {\bf 40}, 659 (1984).)

\bibitem{chiu}C. B. Chiu, J. Pasupathy, and S. J. Wilson, Phys. Rev. D {\bf 33}, 1961 (1986)
; C. B. Chiu, S. L. Wilson, J. Pasupathy, and J. P. Singh, Phys.
Rev. D {\bf 36}, 1451, 1553 (1987).

\bibitem{zsl}Shi-Lin Zhu, W-Y. P. Hwang and Ze-Sen Yang,
Phys. Rev. D {\bf 57}, 1527 (1998); {\sl ibid.} D{\bf 56}, 7273
(1997); Mod. Phys. Lett.A 12, 3027 (1997); Phys. Lett. B 420, 8
(1998); Shi-Lin Zhu, Yuan-Ben Dai, Phys. Rev. D 59, 114015 (1999);
Shi-Lin Zhu, Phys. Rev. D 61, 114019 (2000).

\bibitem{first}V. M. Braun and I. E. Filyanov,
Z. Phys. C {\bf 44}, 157 (1989).

\bibitem{aliev}T.M. Aliev, I. Kanik, M. Savci, Phys. Rev. D 68, 056002 (2003);
T.M. Aliev, A. Ozpineci, M. Savci. Phys. Rev. D 66, 016002
(2002),Erratum-ibid. D 67, (039901) 2003; Phys. Rev. D 65, 096004
(2002); Phys. Rev. D 65, 056008 (2002); Phys. Lett. B 516, 299
(2001); Nucl. Phys. A 678, 443 (2000); T.M. Aliev and A. Ozpineci.
Phys. Rev. D 62, 053012 (2000).

\bibitem{photon}V. M. Braun and I. E. Filyanov, Z. Phys. C 48, 239
(1990);  A. Ali and V. M. Braun, Phys. Lett. B 359, 223 (1995);G.
Eilam, I. Halperin and R. R. Mendel, Phys. Lett. B {\bf 361}, 137
(1995); P. Ball, V. M. Braun and N. Kivel, Nucl. Phys. B {\bf
649}, 263 (2003).


\end{thebibliography}
\end{document}